\documentclass[aps,preprint,showpacs,superscriptaddress]{revtex4}%
\usepackage{amsfonts}
\usepackage{amsmath}
\usepackage{amssymb}
\usepackage{graphicx}%
\begin{document}
\preprint{HEP/123-qed}
\title[Short title for running header]{Effects of magnetic field and transverse anisotropy on full counting
statistics in single-molecule magnet}
\author{Hai-Bin Xue}
\email{xhb98326110@163.com}
\affiliation{Institute of Theoretical Physics, Shanxi University, Taiyuan, Shanxi 030006, China}
\author{Y.-H. Nie}
\email{Nieyh@sxu.edu.cn}
\affiliation{Institute of Theoretical Physics, Shanxi University, Taiyuan, Shanxi 030006, China}
\affiliation{Institute of Solid State Physics, Shanxi Datong University, Datong, 037009, China}
\author{Z.-J. Li}
\affiliation{Institute of Theoretical Physics, Shanxi University, Taiyuan, Shanxi 030006, China}
\author{J.-Q. Liang}
\affiliation{Institute of Theoretical Physics, Shanxi University, Taiyuan, Shanxi 030006, China}

\begin{abstract}
We have theoretically studied the full counting statistics of
electron transport through a single-molecule magnet (SMM) with an
arbitrary angle between the applied magnetic field and the SMM's
easy axis above the sequential tunneling threshold, since the angle
$\theta$ cannot be controlled in present-day SMM experiments. In the
absence of the small transverse anisotropy, when the coupling of the
SMM with the incident-electrode is stronger than that with the
outgoing-electrode, i.e., $\Gamma_{L}/\Gamma _{R}\gg1$, the maximum
peak of shot noise first increases and then decreases with
increasing $\theta$ from $0$ to $0.5\pi$. In particular, the shot
noise can reach up to super-Poissonian value from sub-Poissonian
value when considering the small transverse anisotropy. For
$\Gamma_{L}/\Gamma_{R}\ll1$, the maximum peaks of the shot noise and
skewness can be reduced from a super-Poissonian to a sub-Poissonian
value with increasing $\theta$ from $0$ to $0.5\pi$; the
super-Poissonian behavior of the skewness is more sensitive to the
small $\theta$ than shot noise, which is suppressed when taking into
account the small transverse anisotropy. These characteristics of
shot noise can be qualitatively attributed to the competition
between the fast and slow transport channels. The predictions
regarding of the $\theta$-dependence of high order current cumulants
are very interesting for a better understanding electron transport
through SMM, and will allow for experimental tests in the near
future.

\end{abstract}
\date{\today}
\maketitle

\section{INTRODUCTION}

Electronic transport through an individual single-molecule magnet
(SMM) has attracted intense
experimental\cite{Heersche,Jo,Grose,Zyazin} and
theoretical\cite{Romeike1,Romeike2,Elste,Timm,Xue,Xue02,Leuenberger,Misiorny,LuHZ}
investigation due to its potential application in molecular
spintronics devices\cite{Lapo} and classical\cite{Affronte} and
quantum information processing\cite{Leuenberger1,Lehmann}. The
prototypal SMM is characterized by a large spin ($S>1/2$), easy-axis
anisotropy which defines the preferred z axis in spin space along
which spin is quantized, and transverse anisotropies which allow
tunneling transitions between the molecular eigenstates of
$S_{tot}^{z}$. Since the transverse anisotropy may lead to mixing of
spin eigenstates of $S_{tot}^{z}$, the molecular eigenstates are not
simultaneous eigenstates of $S_{tot}^{z}$. However, these transverse
anisotropy terms are very small compared with the easy-axis
anisotropy so that they can be taken into account by the standard
perturbation calculation. On the other hand, the effect of an
external strong magnetic field on electron transport through the SMM
has also been studied, in which the easy axis of the SMM is usually
assumed to along the direction of the external magnetic field
$\vec{B}$. In the present actual break-junction and electromigration
experiments, however, the angle of the external field with respect
to the easy axis of the SMM is unknown and cannot be
controlled\cite{Heersche,Jo,Grose,Zyazin}. If the angle between the
easy axis and magnetic field is not small, the transverse Zeeman
energy may compare with the easy-axis anisotropy energy. This
implies that the molecular eigenstates are not approximate
eigenstates of the spin component along any axis, which leads to the
failure of the perturbation calculation. In very recent
single-molecule experiment, Zyazin \textit{et al.}\cite{Zyazin}
found that the angle between $\vec{B}$ and the easy axis of the SMM
plays an important role in fitting the theoretical model to
experimental data. Therefore, it is significant to study the effect
of the angle between the SMM's easy axis and magnetic field on
electron transport in the SMM system.

Although the present experimental studies focused on the differential
conductance or average current\cite{Heersche,Jo,Grose,Zyazin}, the full
counting statistics (FCS) of electron transport through single-molecule magnet
or molecular junction has been attracting much theoretical research interests
\cite{Romeike2,Xue,Xue02,Djukic,Imura,Welack,Dong,Aguado} owing to its
allowing one to identify the internal level structure of the transport
system\cite{Xue,Xue02,Belzig} and to access information of electron
correlation that can not be contained in the differential conductance and the
average current\cite{Blanter}. For example, our previous
studies\cite{Xue,Xue02} have shown that the super-Poissonian noise
characteristics of electron transport through the SMM can be employed to
reveal important information of the internal level structure of the SMM and
the left-right asymmetry of the SMM-electrode coupling. In addition, the
frequency-resolved shot noise spectrum of artificial SMM, e.g., a CdTe quantum
dot doped with a single $S=5/2$ Mn spin, can allow one to separately extract
the hole and Mn spin relaxation times via the Dicke effect\cite{Aguado}.
Especially, the FCS may provide the full information about the probability
distribution $P\left(  n,t\right)  $ of transferring $n$ electrons between
electrode and SMM during a time interval $t$. The FCS may be obtained from the
cumulant generating function (CGF) $F\left(  \chi\right)  $ which related to
the probability distribution by\cite{Bagrets}%
\begin{equation}
e^{-F\left(  \chi\right)  }=\sum_{n}P\left(  n,t\right)  e^{in\chi},
\label{CGF}%
\end{equation}
where $\chi$ is the counting field. All cumulants of the current can be
obtained from the CGF by performing derivatives with respect to the counting
field $C_{k}=\left.  -\left(  -i\partial_{\chi}\right)  ^{k}F\left(
\chi\right)  \right\vert _{\chi=0}$. In the long-time limit, the first three
cumulants are directly related to the transport characteristics. For example,
the first-order cumulant (the peak position of the distribution of
transferred-electron number) $C_{1}=\bar{n}$ gives the average current
$\left\langle I\right\rangle =eC_{1}/t$. The zero-frequency shot noise is
related to the second-order cumulant (the peak-width of the distribution)
$S=2e^{2}C_{2}/t=2e^{2}\left(  \overline{n^{2}}-\bar{n}^{2}\right)  /t$. The
third cumulant $C_{3}=\overline{\left(  n-\bar{n}\right)  ^{3}}$ characterizes
the skewness of the distribution. Here, $\overline{\left(  \cdots\right)
}=\sum_{n}\left(  \cdots\right)  P\left(  n,t\right)  $. In general, the shot
noise and the skewness are represented by the Fano factor $F_{2}=C_{2}/C_{1}$
and $F_{3}=C_{3}/C_{1}$, respectively.

In this work, we consider a more universal model of the SMM and investigate
the effect of the angle between the easy axis of the SMM and the applied
magnetic field and the transverse anisotropy on the FCS in SMM. Up to now, the
effects of the angle between the easy axis and magnetic field, and the
transverse anisotropy on the FCS in the present SMM system has not been
studied to the best of our knowledge. We found that although the threshold
bias voltage\ of the sequential tunneling has only a tiny decrease with the
increase of the angle $\theta$ from $0$ to $0.5\pi$,\ the quantum noise
properties of electron transport through SMM is\ not only depend on the
left-right asymmetry of the SMM-electrode coupling, but also the angle
$\theta$ between the easy axis and magnetic field, which can be qualitatively
attributed to the competition between the fast and the slow transport
channels. The paper is organized as follows. In Sec. II, we introduce the SMM
system and outline the procedure to obtain the FCS formalism based on an
effective particle-number-resolved quantum master equation and the
Rayleigh--Schr\"{o}dinger perturbation theory. The numerical results are
discussed in Sec. III, where we discuss the effects of an arbitrary angle
between the easy axis of the SMM and the applied magnetic field, and the
second-order transverse anisotropy on the super-Poissonian noise, and analyze
the occurrence-mechanism of super-Poissonian noise. Finally, in Sec. IV we
summarize the work.

\section{MODEL AND FORMALISM}

A SMM coupled to two metallic electrodes L (left) and R (right) is described
by the Hamiltonian $H_{total}=H_{mol}+H_{leads}+H_{T}$. We assume that the
SMM-electrode coupling is sufficiently weak so that the electron transport is
dominated by sequential tunneling. The SMM Hamiltonian is given by
\begin{align}
H_{mol}  &  =(\varepsilon_{d}-eV_{g})\hat{n}+\frac{U}{2}\hat{n}(\hat
{n}-1)-J\,\vec{s}\cdot\vec{S}\nonumber\\
&  -K_{1}(S_{z})^{2}+K_{2}\left(  S_{+}^{2}+S_{-}^{2}\right)  -\vec{B}%
\cdot\left(  \vec{s}+\vec{S}\right)  . \label{model}%
\end{align}
Here, the first two terms depict the lowest unoccupied molecular orbital
(LUMO), $\hat{n}\equiv d_{\uparrow}^{\dag}d_{\uparrow}+d_{\downarrow}^{\dag
}d_{\downarrow}$ is the number operator of the electron in the molecule, where
$d_{\sigma}^{\dag}$ ($d_{\sigma}$) creates (annihilates) an electron with spin
$\sigma$ and energy $\varepsilon_{d}$ (which can be tuned by a gate voltage
$V_{g}$). $U$ is the Coulomb repulsion between two electrons in the LUMO. The
third term describes the exchange coupling between\ electron spin in the LUMO
and the giant spin, the electronic spin operator $\vec{s}\equiv\sum
_{\sigma\sigma^{\prime}}d_{\sigma}^{\dag}\left(  \vec{\sigma}_{\sigma
\sigma^{\prime}}\right)  d_{\sigma^{\prime}}$ with $\vec{\sigma}\equiv$
$(\sigma_{x},\sigma_{y},\sigma_{z})$ being the vector of Pauli matrices. The
forth and fifth terms are the anisotropy energies of the SMM, where $K_{1}$
describes the easy-axis anisotropy and $K_{2}$ the transverse anisotropy. The
last term denotes Zeeman splitting, where $g\mu_{B}$ has been absorbed into
$\vec{B}$. In a general case, since the transverse anisotropy and the magnetic
field terms do not commute with the easy-axis anisotropy term, $S_{tot}^{z}$
$\left(  =s^{z}+S^{z}\right)  $ is not conserved and the SMM eigenstates of
$H_{mol}$ are not simultaneous eigenstates of $S_{tot}^{z}$. On the other
hand, if the external magnetic field $\vec{B}$ is applied along the easy-axis,
in the absence of the transverse anisotropy the eigenvalue $m$ of $S_{tot}%
^{z}$ is a good quantum number, hence allowing us to numerically
diagonalize the molecular Hamiltonian $H_{mol}$ in the basis
represented by the eigenvalue $m$ of $S_{tot}^{z}$ and the
corresponding occupation number $n$ of the LUMO level, i.e.,
$\left\{  \left\vert 0,m\right\rangle ,\left\vert \downarrow
\right\rangle \left\vert m-\frac{1}{2}\right\rangle ,\left\vert
\uparrow \right\rangle \left\vert m+\frac{1}{2}\right\rangle
,\left\vert 2,m\right\rangle \right\}  $, where $m\in\left[
-S,S\right]  $.

The relaxation in the electrodes is assumed to be sufficiently fast so that
their electron distributions can be described by equilibrium Fermi functions.
The electrodes are modeled as non-interacting Fermi gases and the
corresponding Hamiltonian%
\begin{equation}
H_{Leads}=\sum_{\alpha\mathbf{k}\sigma}\varepsilon_{\alpha\mathbf{k}\sigma
}a_{\alpha\mathbf{k}\sigma}^{\dag}a_{\alpha\mathbf{k}\sigma}, \label{Leads}%
\end{equation}
where $a_{\alpha\mathbf{k}\sigma}^{\dag}$ ($a_{\alpha\mathbf{k}\sigma}$)
creates (annihilates) an electron with energy $\varepsilon_{\alpha
\mathbf{k}\sigma}$, momentum $\mathbf{k}$, and spin $\sigma$ in $\alpha$
($\alpha=L,R$) electrode. The tunneling between the LUMO and the electrodes is
described by
\begin{equation}
H_{T}=\sum_{\alpha\mathbf{k}\sigma}\left(  t_{\alpha}a_{\alpha\mathbf{k}%
\sigma}^{\dag}d_{\sigma}+H.c.\right)  . \label{tunneling}%
\end{equation}

In sequential tunneling regime, the transitions are well described by quantum
master equation of a reduced density matrix spanned by the eigenstates of the
SMM. The detailed derivation of the FCS based on the particle-number-resolved
quantum master equation can be found in Refs. \cite{Li1,Li2,WangSK}, and here,
we only give the main results. Under the second order Born approximation and
Markovian approximation, the particle-number-resolved quantum master equation
for the reduced density matrix is given by
\begin{equation}
\dot{\rho}^{\left(  n\right)  }\left(  t\right)  =-i\mathcal{L}\rho^{\left(
n\right)  }\left(  t\right)  -\frac{1}{2}\mathcal{R}\rho^{\left(  n\right)
}\left(  t\right)  , \label{Master1}%
\end{equation}
with%
\begin{align}
\mathcal{R}\rho^{\left(  n\right)  }\left(  t\right)   &  =%
{\displaystyle\sum\limits_{\mu=\uparrow,\downarrow}}
\left[  d_{\mu}^{\dagger}A_{\mu}^{\left(  -\right)  }\rho^{\left(  n\right)
}\left(  t\right)  +\rho^{\left(  n\right)  }\left(  t\right)  A_{\mu
}^{\left(  +\right)  }d_{\mu}^{\dagger}\right. \nonumber\\
&  \left.  -A_{L\mu}^{\left(  -\right)  }\rho^{\left(  n\right)  }\left(
t\right)  d_{\mu}^{\dagger}-d_{\mu}^{\dagger}\rho^{\left(  n\right)  }\left(
t\right)  A_{L\mu}^{\left(  +\right)  }\right. \nonumber\\
&  \left.  -A_{R\mu}^{\left(  -\right)  }\rho^{\left(  n-1\right)  }\left(
t\right)  d_{\mu}^{\dagger}-d_{\mu}^{\dagger}\rho^{\left(  n+1\right)
}\left(  t\right)  A_{R\mu}^{\left(  +\right)  }\right]  +H.c.,
\label{Master2}%
\end{align}
where $A_{\mu}^{\left(  \pm\right)  }=\sum_{\alpha=L,R}A_{\alpha\mu}^{\left(
\pm\right)  }$, $A_{\alpha\mu}^{\left(  \pm\right)  }=\Gamma_{\alpha}%
n_{\alpha}^{\pm}\left(  -\mathcal{L}\right)  d_{\mu}$, $n_{\alpha}%
^{+}=f_{\alpha},n_{\alpha}^{-}=1-f_{\alpha}$ ($f_{\alpha}$ is the Fermi
function of the electrode $\alpha$), and $\Gamma_{\alpha=L,R}=2\pi
g_{\alpha=L,R}\left\vert t_{\alpha=L,R}\right\vert ^{2}$. Liouvillian
superoperator $\mathcal{L}$ is defined as $\mathcal{L}\left(  \cdots\right)
=\left[  H_{mol},\left(  \cdots\right)  \right]  $, and $g_{\alpha=L,R}$ are
the density of states of the metallic electrodes. $\rho^{\left(  n\right)
}\left(  t\right)  $ describes the reduced density matrix of the SMM
conditioned by the electron numbers tunneling through the right junction up to
time $t$. Throughout this work, we set $e\equiv\hbar=1$. Here, the validity of
the Markovian approximation deserves some discussions. For the case of
sequential tunneling, the Markovian approximation is valid when the system
conductance is small compared to the quantum conductance\cite{Timm-MP}, i.e.,
$I/V\ll$ $e^{2}/\left(  2\pi\hbar\right)  =1/\left(  2\pi\right)  $, here we
have utilized $e\equiv\hbar=1$. In the present SMM system, the value of $I/V$
is of the order of $10^{-3}\ll1/\left(  2\pi\right)  $. This means that the
typical time between two tunneling events is $\tau_{0}=e/I\gg2\pi\hbar/\left(
eV\right)  =\tau_{leads}$, i.e., the SMM dynamics is indeed much slower than
the decay of lead correlations, thus, the Markovian approximation is well
justified\cite{Timm-MP}. The CGF connects with the particle-number-resolved
density matrix by defining $S\left(  \chi,t\right)  =\sum_{n}\rho^{\left(
n\right)  }\left(  t\right)  e^{in\chi}$. Evidently, we have $e^{-F\left(
\chi\right)  }=$Tr$\left[  S\left(  \chi,t\right)  \right]  $, where the trace
is over the eigenstates of the SMM. Since Eq. (\ref{Master1}) has the
following form
\begin{equation}
\dot{\rho}^{\left(  n\right)  }=A\rho^{\left(  n\right)  }+C\rho^{\left(
n+1\right)  }+D\rho^{\left(  n-1\right)  }, \label{formalmaster}%
\end{equation}
then $S\left(  \chi,t\right)  $ satisfies
\begin{equation}
\dot{S}=AS+e^{-i\chi}CS+e^{i\chi}DS\equiv L_{\chi}S, \label{formalmaster1}%
\end{equation}
where the specific form of $L_{\chi}$ can be obtained by performing a discrete
Fourier transformation to the matrix element of Eq. (\ref{Master1}). Here, the
master equation contains off-diagonal matrix elements $\rho_{mn}^{SMM}$, which
corresponds to superpositions between molecular eigenstates $\left\vert
m\right\rangle $ and $\left\vert n\right\rangle $. In fact, since the presence
of noncommuting Zeeman and transverse anisotropy terms in the SMM Hamiltonian,
any two eigenstates differ in the spin expectation value $\left\langle
S_{tol}\right\rangle $, which leads to different long-range (dipole) fields.
Thus the unavoidable interactions between the SMM and many degrees of freedom
in the environment (e.g., electron spins) lead to rapid decay of
superpositions of these eigenstates and thus of $\rho_{mn}^{SMM}%
$\cite{Timm,Timm-MP,Zurek}. As a result, in the following calculation the
off-diagonal matrix elements can be neglected, and it is sufficient to
consider the diagonal components of $\rho^{SMM}$.

In the low frequency limit, the counting time ($i.e.$, the time of
measurement) is much longer than the time of tunneling through the SMM. In
this case, $F\left(  \chi\right)  $ is given
by\cite{Bagrets,Flindt,Kie-lich,Groth}%
\begin{equation}
F\left(  \chi\right)  =-\lambda_{1}\left(  \chi\right)  t, \label{CGFformal}%
\end{equation}
where $\lambda_{1}\left(  \chi\right)  $ is the eigenvalue of $L_{\chi}$ which
goes to zero for $\chi\rightarrow0$. According to the definition of the
cumulants one can express $\lambda_{1}\left(  \chi\right)  $\ as
\begin{equation}
\lambda_{1}\left(  \chi\right)  =\sum_{k=1}^{\infty}\frac{C_{k}}{t}%
\frac{\left(  i\chi\right)  ^{k}}{k!}. \label{Lambda}%
\end{equation}
Low order cumulants can be calculated by the Rayleigh--Schr\"{o}dinger
perturbation theory in the counting parameter $\chi$. In order to calculate
the first three current cumulants we expand $L_{\chi}$ to third order in
$\chi$%
\begin{equation}
L_{\chi}=L_{0}+L_{1}\chi+\frac{1}{2!}L_{2}\chi^{2}+\frac{1}{3!}L_{3}\chi
^{3}+\cdots. \label{matirxL}%
\end{equation}
Along the lines of Ref. \cite{Flindt}, we define the two projectors
$\mathcal{P}=\mathcal{P}^{2}=\left\vert \left.  0\right\rangle \right\rangle
\left\langle \left\langle \tilde{0}\right.  \right\vert $ and $\mathcal{Q}%
=\mathcal{Q}^{2}=1-\mathcal{P}$, obeying the relations $\mathcal{P}L_{0}%
=L_{0}\mathcal{P}=0$ and $\mathcal{Q}L_{0}=L_{0}\mathcal{Q=}L_{0}$. Here,
$\left\vert \left.  0\right\rangle \right\rangle $ being the steady state
$\rho^{stat}$ is the right eigenvectors of $L_{0}$, i.e., $L_{0}\left\vert
\left.  0\right\rangle \right\rangle =0$, and $\left\langle \left\langle
\tilde{0}\right.  \right\vert \equiv\hat{1}$ is the corresponding left
eigenvectors. In view of $L_{0}$ is regular, we can also introduce the
pseudoinverse according to $R=\mathcal{Q}L_{0}^{-1}\mathcal{Q}$, which is
well-defined because the inversion is performed only in the subspace spanned
by $\mathcal{Q}$. After a careful calculation, $\lambda_{1}\left(
\chi\right)  $ is given by%

\begin{align}
\lambda_{1}\left(  \chi\right)   &  =\left\langle \left\langle \tilde
{0}\right.  \right\vert L_{1}\left\vert \left.  0\right\rangle \right\rangle
\chi\nonumber\\
&  +\frac{1}{2!}\left[  \left\langle \left\langle \tilde{0}\right.
\right\vert L_{2}\left\vert \left.  0\right\rangle \right\rangle
-2\left\langle \left\langle \tilde{0}\right.  \right\vert L_{1}RL_{1}%
\left\vert \left.  0\right\rangle \right\rangle \right]  \chi^{2}\nonumber\\
&  +\frac{1}{3!}\left[  \left\langle \left\langle \tilde{0}\right.
\right\vert L_{3}\left\vert \left.  0\right\rangle \right\rangle
-3\left\langle \left\langle \tilde{0}\right.  \right\vert \left(  L_{2}%
RL_{1}+L_{1}RL_{2}\right)  \left\vert \left.  0\right\rangle \right\rangle
\right. \nonumber\\
&  \left.  -6\left\langle \left\langle \tilde{0}\right.  \right\vert
L_{1}R\left(  RL_{1}P-L_{1}R\right)  L_{1}\left\vert \left.  0\right\rangle
\right\rangle \right]  \chi^{3}+\cdots. \label{matrixLambda}%
\end{align}
From Eqs. (\ref{Lambda}) and (\ref{matrixLambda}) we can identify the first
three current cumulants:%
\begin{equation}
C_{1}/t=\left\langle \left\langle \tilde{0}\right.  \right\vert L_{1}%
\left\vert \left.  0\right\rangle \right\rangle /i, \label{current}%
\end{equation}%
\begin{equation}
C_{2}/t=\left[  \left\langle \left\langle \tilde{0}\right.  \right\vert
L_{2}\left\vert \left.  0\right\rangle \right\rangle -2\left\langle
\left\langle \tilde{0}\right.  \right\vert L_{1}RL_{1}\left\vert \left.
0\right\rangle \right\rangle \right]  /i^{2}, \label{shot noise}%
\end{equation}%
\begin{align}
&  C_{3}/t=\left[  \left\langle \left\langle \tilde{0}\right.  \right\vert
L_{3}\left\vert \left.  0\right\rangle \right\rangle -3\left\langle
\left\langle \tilde{0}\right.  \right\vert \left(  L_{2}RL_{1}+L_{1}%
RL_{2}\right)  \left\vert \left.  0\right\rangle \right\rangle \right.
,\nonumber\\
&  \left.  -6\left\langle \left\langle \tilde{0}\right.  \right\vert
L_{1}R\left(  RL_{1}P-L_{1}R\right)  L_{1}\left\vert \left.  0\right\rangle
\right\rangle \right]  /i^{3}. \label{skewness}%
\end{align}
The three equations above are the starting point of the calculation in following.

\section{NUMERICAL RESULTS AND DISCUSSION}

We now study the effects of the angle of the external field with respect to
the easy axis of the SMM and the transverse anisotropy on the FCS of
electronic transport through the SMM weakly coupled to two metallic
electrodes. We assume the bias voltage ($V_{b}=\mu_{L}-\mu_{R}$) is
symmetrically entirely dropped at the SMM-electrode tunnel junctions, which
implies that the levels of the SMM are independent of the applied bias voltage
even if the couplings are not symmetric. Since our previous work\cite{Xue02}
has studied the effect of Coulomb interaction $U$ on FCS in the SMM in the
absence of the transverse magnetic fields and transverse anisotropy, we here
take a fixed value of $U$. The parameters of the SMM are chosen as\cite{Timm}
$S=2$, $\varepsilon_{d}=200\Gamma$, $U=100\Gamma$, $J=100\Gamma$,
$K_{1}=40\Gamma$\ and $\left\vert \vec{B}\right\vert =80\Gamma$, where
$\Gamma$\ is the typical tunneling rate of electrons between the SMM and the
electrode. In the present work, we only study the transport above the
sequential tunneling threshold, i.e., $V_{b}>2\epsilon_{se}$, where
$\epsilon_{se}$\ is the energy difference between the ground state with charge
$N$\ and the first excited states $N-1$\cite{Aghassi1}. In this regime, the
inelastic sequential tunneling process is dominant, thus electrons have
sufficient energy to overcome the Coulomb blockade and tunnel sequentially
through the SMM. Here, it should be noted that since in the Coulomb blockade
regime the current is exponentially suppressed and the electron transport is
dominated by cotunneling, when taking into account cotunneling the normalized
second and third moments will deviate from the results obtained by only
sequential tunneling\cite{Thielmann}. In this paper, we put emphasis on the
effects of the angle between $\vec{B}$ and the easy-axis of the SMM, and the
transverse anisotropy on super-Poissonian noise for large left-right asymmetry
of the SMM-electrode coupling.

Since the transverse anisotropy and the transverse component of the applied
magnetic field can lead to mixing of spin eigenstates of $S_{tot}^{z}$, the
transitions, which are inhibited due to spin selection rules in the absence of
a transverse field and transverse anisotropy, may occur. In order to show
explicitly the effect of the angle between $\vec{B}$ and the easy axis on
electron transport, we first neglect the small transverse anisotropy. In this
case, the applied magnetic field $\vec{B}$ may be assumed to lie in the $xz$
plane because of the rotational symmetry of $H_{mol}$. Moreover, it is helpful
to analyze the selection rules for the occurrence of the sequential tunneling.
In the absence of the transverse fields and the transverse anisotropy, the
eigenvalue $m$ of $S_{tot}^{z}$ is a good quantum number and the sequential
tunneling requires a change of the electron number by $\Delta n=\pm1$, and the
magnetic quantum number by $\Delta m=\pm1/2$. But for the present case, the
only selection rule $\Delta n=\pm1$ is still valid, which means that arbitrary
two states satisfying $\Delta n=\pm1$ can do sequential tunneling. For local
large spin $S$, there are $2S+1$ empty molecular states with $n=0$ in the
LUMO, $2\left(  2S+1\right)  $ singly-occupied molecular states with $n=1$ and
$2S+1$ doubly-occupied states with $n=2$. Therefore, there are $4\left(
2S+1\right)  ^{2}$ transitions, namely, $2\left(  2S+1\right)  ^{2}$ between
molecular states with $n=0$ and $n=1$, and $2\left(  2S+1\right)  ^{2}$
between molecular states with $n=1$ and $n=2$, which leads to a much more
complex electron transport channels than the case without the transverse
anisotropy and transverse field. For this reason, we focus on studying the
dependence of the maximum noise values on the angle $\theta$ in sequential
tunneling regime.

When the coupling of the SMM with the left electrode is stronger
than that with the right electrode, i.e.,
$\Gamma_{L}/\Gamma_{R}\gg1$, here we choose
$\Gamma_{L}/\Gamma_{R}=10$. Figures 1(a)-(c) show the average
current, shot noise and skewness as a function of the bias voltage
for $\theta=0$, $0.1\pi$, $0.2\pi$, $0.3\pi$, $0.4\pi$, $0.5\pi$.
Since the FCS for $\theta=\theta_{0}$ has the same
bias-voltage-dependence as that for $\theta=\pi-\theta_{0}$, which
arises from the symmetry of the SMM Hamiltonian, we restrict our
discussion to the case of $\theta\in\left[  0,0.5\pi\right]  $. With
increasing $\theta$, the corresponding sequential tunneling
threshold bias voltage has a tiny decrease and reach their minimums
when $\theta$ increases to $0.5\pi$, see Fig. 1(a); but quantum
noise obviously depends on the angle $\theta$, see Fig. 1(b) and
(c). The maximum peak of shot noise firstly increases and then
decreases with increasing the angle from $0$ to $0.5\pi$. This
characteristics of the shot noise can be understood with the help of
the dynamic competition between effective fast and slow transport
channels\cite{WangSK,Aghassi1,Safonov,Djuric,Aghassi2,Aguado}. The
molecular
channel current is given by\cite{Timm,Xue}%
\begin{align}
&  I_{\left\vert n,i\right\rangle \longrightarrow\left\vert n-1,j\right\rangle
}\nonumber\\
&  =C_{\left\vert n-1,j\right\rangle ,\left\vert n,i\right\rangle }\Gamma
_{R}n_{R}^{-}\left(  \epsilon_{\left\vert n,i\right\rangle }-\epsilon
_{\left\vert n-1,j\right\rangle }-\mu_{R}\right)  P_{\left\vert
n,i\right\rangle },\label{channeladd}\\
&  I_{\left\vert n-1,j\right\rangle \longrightarrow\left\vert n,i\right\rangle
}\nonumber\\
&  =-C_{\left\vert n-1,i\right\rangle ,\left\vert n,j\right\rangle
}\Gamma _{R}n_{R}^{+}\left(  \epsilon_{\left\vert n,i\right\rangle
}-\epsilon _{\left\vert n-1,j\right\rangle }-\mu_{R}\right)
P_{\left\vert
n-1,j\right\rangle }. \label{channelsub}%
\end{align}
Here $C_{\left\vert n-1,j\right\rangle ,\left\vert n,i\right\rangle
}=\left\vert \left\langle n-1,j\right\vert d_{\sigma}\left\vert
n,i\right\rangle \right\vert ^{2}$ is a constant which related to the two
molecular states but independent of the bias voltage, where $\left\vert
n,i\right\rangle $ ($i=01$, $02$, $03$, $04$, $05$, for $S=2$) denote the
eigenstates of the molecule with $n$ electrons tunneling into the molecule,
which are arranged in an ascending order of their eigenvalues $\epsilon
_{\left\vert n,i\right\rangle }$. $P_{\left\vert n,i\right\rangle }$ is the
occupied probability of the state $\left\vert n,i\right\rangle $. Since the
maximum value of shot noise appears at a large bias voltage, the Fermi
function $f_{R}\left(  \epsilon_{\left\vert n,i\right\rangle }-\epsilon
_{\left\vert n-1,j\right\rangle }-\mu_{R}\right)  $ changes very slowly with
increasing bias voltage, i.e., $f_{R}\left(  \epsilon_{\left\vert
n,i\right\rangle }-\epsilon_{\left\vert n-1,j\right\rangle }-\mu_{R}\right)
\simeq0$. Thus the molecular channel currents $I_{\left\vert n,i\right\rangle
\longrightarrow\left\vert n-1,j\right\rangle }$ are mainly determined by the
probability distribution$\ P_{\left\vert n,i\right\rangle }$, and
$I_{\left\vert n-1,j\right\rangle \longrightarrow\left\vert n,i\right\rangle
}\simeq0$. In the presence of the transverse field and the transverse
anisotropy, since the transitions between the molecular eigenstates are not
restricted by the selection rule $\Delta m=\pm1/2$, the possible transport
channels are $4\left(  2S+1\right)  ^{2}$, for example, there are a hundred
transport channels for $S=2$. Therefore, it is unpractical to give all the
channel currents. In order to give a qualitative explanation for the effect of
the angle $\theta$ on the shot noise, we plot the occupied probability of the
five main molecular eigenstates as a function of bias voltage $V_{b}$ for
$\theta=0$, $0.1\pi$, $0.2\pi$, $0.3\pi$, $0.4\pi$, $0.5\pi$ in Fig. 2. For
the case of $K_{2}=0$, the increase (or decrease) of the probability of the
molecular eigenstate with high occupancy is always accompanied by the decrease
(or increase) of the probability of the molecular eigenstates with the low
occupancy. It is important that the active competition between the fast and
slow transport channels depends on the angle $\theta$. The competition between
the probability of the five main molecular eigenstates for $\theta=0.2\pi$,
$0.3\pi$ and $0.4\pi$, as shown in Fig. 2, is stronger than that for
$\theta=0$ and $0.5\pi$. Thus, the corresponding transport channel currents
can form the so-called effective fast-and-slow transport channels, which leads
to the maximum value of shot noise for $\theta=0.2\pi$, $0.3\pi$ and $0.4\pi$
are larger than that for $\theta=0$ and $0.5\pi$. In addition, it is
interesting that some certain angles $\theta$ (e.g., $\theta=0.1\pi$, $0.4\pi
$, $0.5\pi$) may decrease the maximum super-Poissonian value of the skewness
$F_{3}>1$ to sub-Poissonian value of $F_{3}<1$ although the angle (e.g.,
$\theta=0.2\pi$) can also increase the maximum skewness value, see Fig. 1(c).
In the simultaneity presence of the transverse field and the small transverse
anisotropy, the active competition is further strengthened, so that the
maximum shot noise value, which is sub-Poissonian value for $\theta=0$, may be
enhanced to super-Poissonian value for some certain angles, e.g.,
$\theta=0.2\pi$, $0.3\pi$ and $0.4\pi$, see Fig. 1(e).

Compared to the case of $\Gamma_{L}/\Gamma_{R}\gg1$, for $\Gamma_{L}%
/\Gamma_{R}\ll1$ the maximum shot noise and the skewness peaks are suppressed
with increasing the angle $\theta$. Fig. 3(a)-(c) show the average current,
shot noise and skewness as a function of the bias voltage for $\theta=0$,
$0.1\pi$, $0.2\pi$, $0.3\pi$, $0.4\pi$, $0.5\pi$ at $\Gamma_{L}/\Gamma
_{R}=0.1$. In this situation, the sequential tunneling threshold
$\epsilon_{se}$ has the same characteristics as the case of $\Gamma_{L}%
/\Gamma_{R}\gg1$. Apart from this feature, another important finding is that
with increasing the angle $\theta$ from $0$ to $0.5\pi$, the maximum peaks of
the shot noise and skewness can be reduced from super-Poissonian to
sub-Poissonian value in the absence of the transverse anisotropy, see Fig.
3(b) and (c). Especially for the skewness, its super-Poissonian behavior seems
more sensitive to the small $\theta$ than shot noise, see Fig. 3(c). The shot
noise characteristics can also be understood in terms of the so-called fast
and slow transport channels mechanism. Fig. 4 shows the occupancy probability
of the five main singly-occupied molecular eigenstates as a function of bias
voltage $V_{b}$ for various values of the angle, which determine corresponding
transport channel currents. With increasing the angle $\theta$ from $0$ to
$0.5\pi$, the competition between the probabilities of the eigenstates with
high occupancy and the eigenstates with the low occupancy is gradually
weakened, see Fig. 4. This means that the active competition between the
fast-and-slow channel currents is gradually suppressed, thus leading to the
maximum super-Poissonian value of shot noise is reduced even to sub-Poissonian
value. Moreover, the small transverse anisotropy for $\Gamma_{L}/\Gamma_{R}%
\ll1$ also suppresses the maximum peaks of the shot noise and the
skewness, and thus effaces the sensitivity of the maximum peaks of
the shot noise and the skewness to the small $\theta$, see Fig. 3(e)
and (f).

\section{CONCLUSIONS}

We have studied the FCS of electron transport through a SMM with an arbitrary
angle between the external magnetic field and the easy axis of the SMM above
the sequential tunneling threshold. Since the presence of the transverse field
and the transverse anisotropy leads to mixing of the spin eigenstates of
$S_{tot}^{z}$, the spin selection rule $\Delta m=\pm1/2$ for sequential
tunneling processes are no longer applied to our model, as a result, there are
$4\left(  2S+1\right)  ^{2}$ transport channels participating in the electron
transport. Therefore, the angle has a complex impact on the FCS. To facilitate
the discussion of the origin of the shot noise, we put special emphasis on the
dependence of the maximum noise on the angle of external magnetic field for
strong asymmetric coupling to the two electrodes. For the case of $\Gamma
_{L}/\Gamma_{R}\gg1$, the maximum peak of the shot noise firstly increase and
then decrease with increasing $\theta$ from $0$ to $0.5\pi$. In particular,
the shot noise is further enhanced and even reaches super-Poissonian value
when considering the small transverse anisotropy. For the case of $\Gamma
_{L}/\Gamma_{R}\ll1$, the maximum peaks of the shot noise and skewness can be
reduced from super-Poissonian to sub-Poissonian value with increasing the
angle $\theta$ from $0$ to $0.5\pi$. Especially for the skewness, its
super-Poissonian behavior seems more sensitive to the small angle $\theta$
than shot noise, but this feature is suppressed when taking into account the
small transverse anisotropy. These characteristics of shot noise can be
understood as a result of the active competition between the fast and slow
transport channels. The predictions regarding the high order current cumulants
are very interesting for better understanding electron transport through
individual single-molecule magnet, and the $\theta$-dependence of the FCS can
be helpful in understanding the experimental results because the angle
$\theta$ is difficult to be controlled experimentally.

\section{ACKNOWLEDGMENTS}

This work was supported by the Graduate Outstanding Innovation Item of Shanxi
Province (Grant No. 20103001), the National Nature Science Foundation of China
(Grant No. 10774094, No. 10775091, No. 10974124 and No. 11075099) and the
Shanxi Nature Science Foundation of China (Grant No. 2009011001-1 and No. 2008011001-2).

\newpage

\begin{figure}[t]
\begin{center}
\includegraphics[height=12cm,width=16cm]{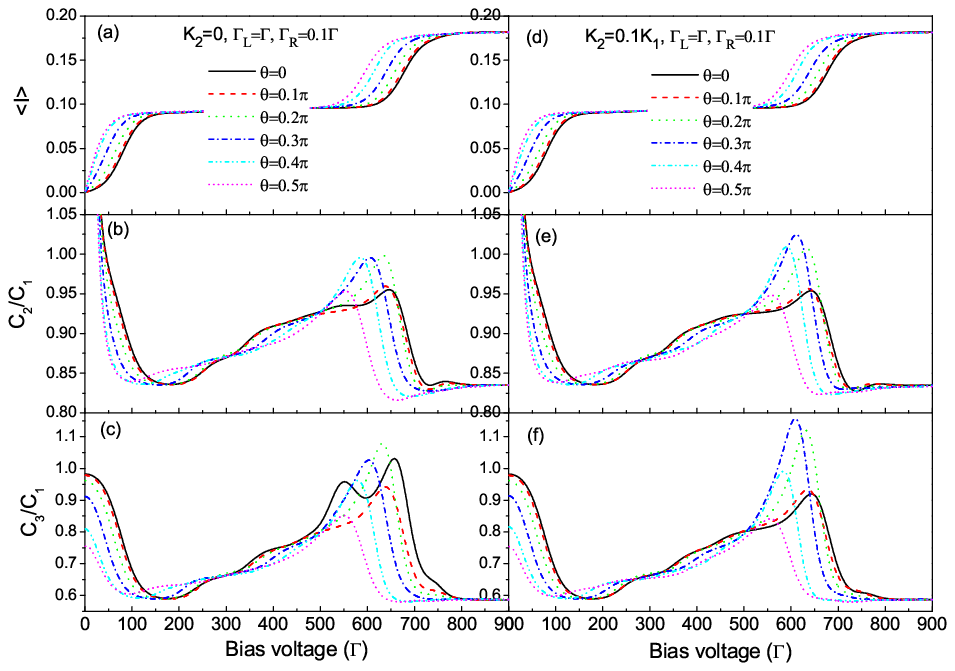}
\end{center}
\caption{(Colour online) The average currant, shot noise and
skewness versus bias voltage for different angles of external
magnetic field with $\Gamma _{L}/\Gamma_{R}=10$. (a), (b) and (c)
for $K_{2}=0$, (d), (e) and (f) for $K_{2}=0.1K_{1}$. The molecular
parameters: $S=2$, $\varepsilon_{d}=200\Gamma $, $U=100\Gamma$,
$J=100\Gamma$, $K_{1}=40\Gamma$, $\left\vert \vec{B}\right\vert
=80\Gamma$
and $k_{B}T=10\Gamma$.}%
\end{figure}

\begin{figure}[t]
\begin{center}
\includegraphics[height=12cm,width=16cm]{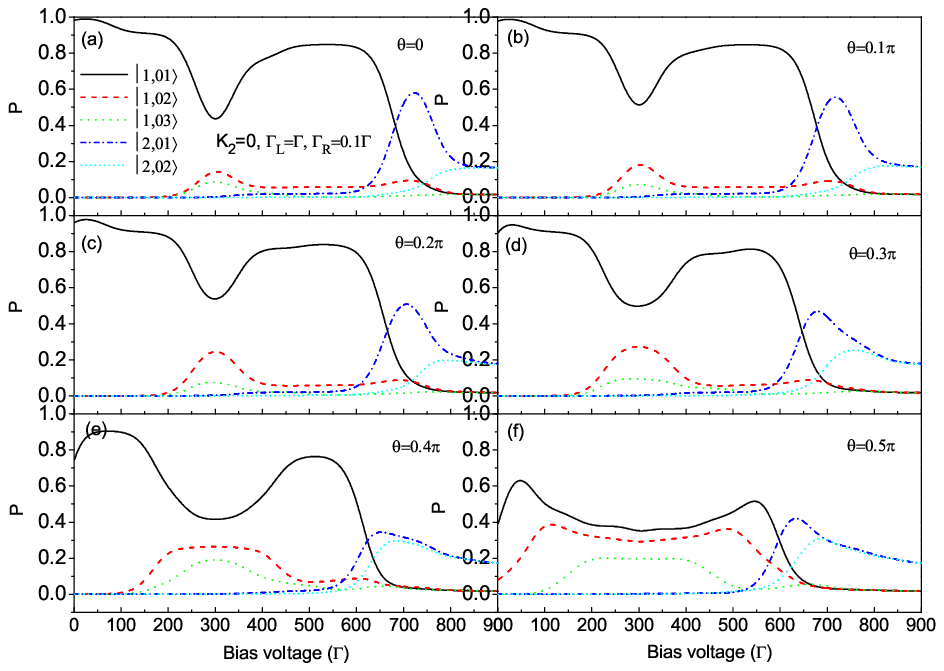}
\end{center}
\caption{(Colour online) The probability distribution of molecular
eigenstates versus bias voltage for different angles of external
magnetic field with $\Gamma_{L}/\Gamma_{R}=10$ and $K_{2}=0$. The
molecular parameters are the same as in Fig. 1.}%
\end{figure}

\begin{figure}[t]
\begin{center}
\includegraphics[height=12cm,width=16cm]{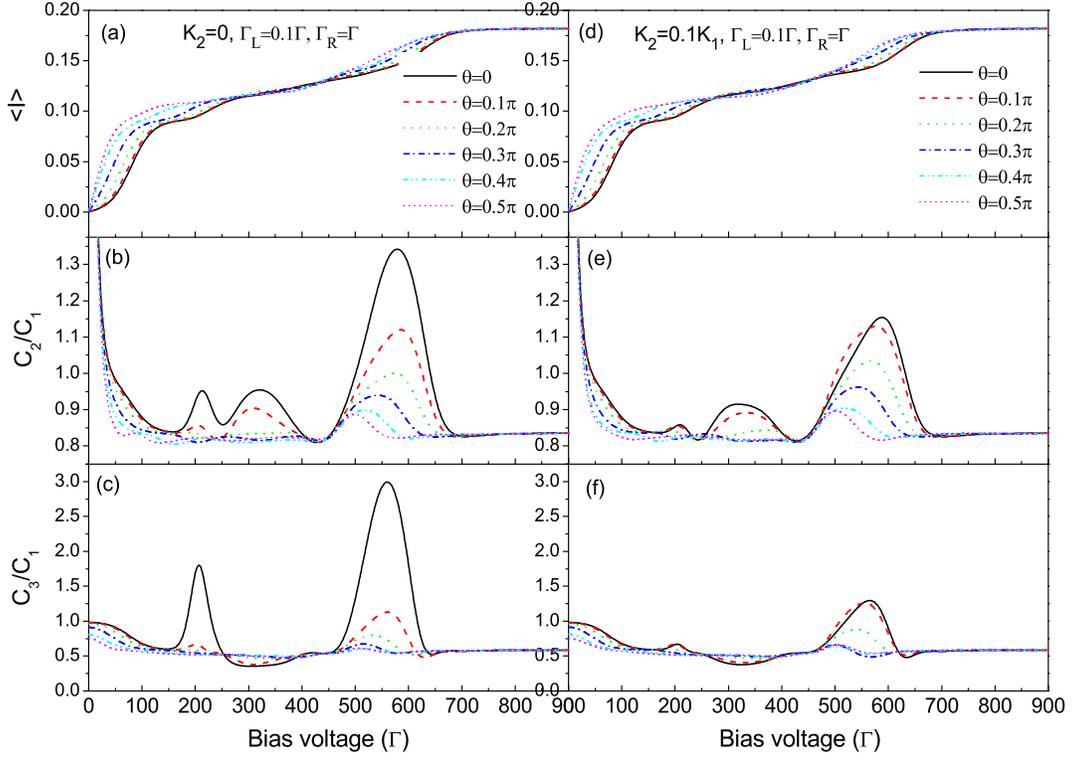}
\end{center}
\caption{(Colour online) The average currant, shot noise and
skewness versus bias voltage for different angles of external
magnetic field with $\Gamma _{L}/\Gamma_{R}=0.1$. (a), (b) and (c)
for $K_{2}=0$, (d), (e) and (f) for $K_{2}=0.1K_{1}$.
The molecular parameters are the same as in Fig. 1.}%
\end{figure}

\begin{figure}[t]
\begin{center}
\includegraphics[height=12cm,width=16cm]{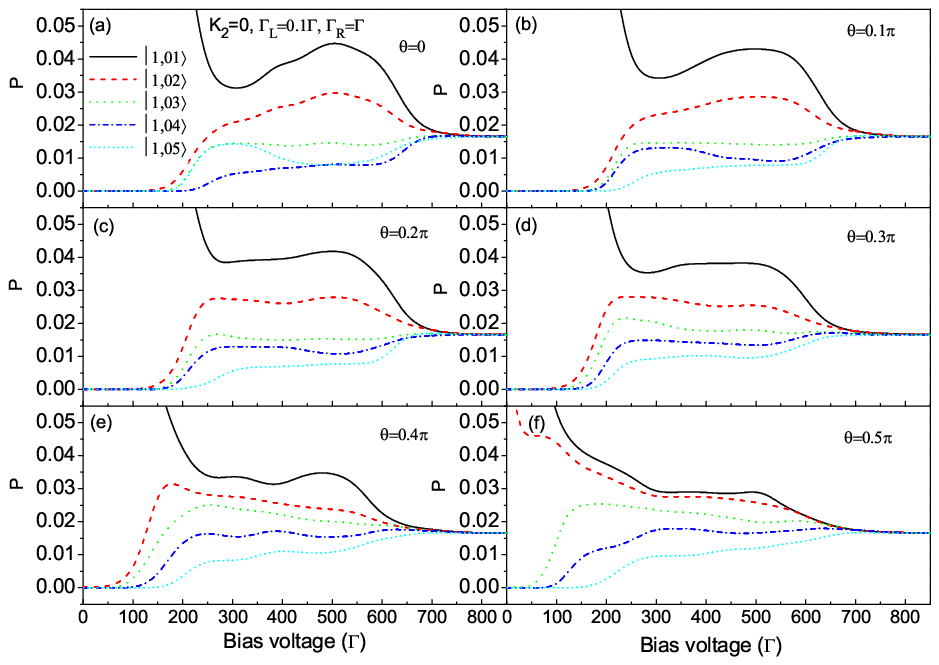}
\end{center}
\caption{(Colour online) The probability distribution of molecular
eigenstates versus bias voltage for different angles of external
magnetic field with $\Gamma_{L}/\Gamma_{R}=0.1$ and $K_{2}=0$. The
molecular parameters are the same as in Fig. 1.}%
\end{figure}

\end{document}